\documentclass[pre,a4paper,oneside,twocolumn,nofootinbib]{revtex4}

\usepackage{color} 
\usepackage[normalem]{ulem} %

\usepackage{graphicx}
\usepackage{hyperref}
\usepackage{amsmath}
\usepackage{amssymb}

\begin{document}

\title{Analytical solution for the long- and short-range every-pair-interactions model}

\author{Fabiano L.\ Ribeiro [0000-0002-2719-6061]}
\email{fribeiro@ufla.br}
\affiliation{Department of Physics (DFI), Federal University of Lavras (UFLA), Lavras MG, Brazil}

\author{Yunfei Li [0000-0001-7023-4599]}
\affiliation{Potsdam Institute for Climate Impact Research -- PIK, P.O.\ Box 601203, 14412 Potsdam, Germany}

\author{Stefan Born [0000-0001-7838-9157]}
\affiliation{Technische Universit\"at Berlin, Chair of Bioprocess Engineering and
Institute of Mathematics, Strasse des 17.\ Juni 135, 10623 Berlin, Germany}

\author{Diego Rybski [0000-0001-6125-7705]}
\email{ca-dr@rybski.de}
\affiliation{Potsdam Institute for Climate Impact Research -- PIK, Member of Leibniz Association, P.O.\ Box 601203, 14412 Potsdam, Germany}
\affiliation{Complexity Science Hub Vienna, Josefst\"adterstrasse 39, A-1090 Vienna, Austria}

\date{\today}

\begin{abstract}
Many physical, biological, and social systems exhibit emergent properties that arise from the interactions between their components (cells). In this study,  we systematically treat every-pair interactions (a) that exhibit power-law dependence on the Euclidean distance  and (b) act in structures that can be characterized using fractal geometry.
We analytically derive the mean interaction field of the cells and find that (i) in a long-range interaction regime, the mean interaction field increases following a power law with the size of the system, (ii) in a short-range interaction regime, the field saturates, and (iii) in the intermediate range
 it follows a logarithmic behaviour.
To validate our analytical solution, we perform numerical simulations. In the case of short-range interactions, we observe that discreteness significantly impacts the continuum approximation used in the derivation, leading to incorrect asymptotic behaviour in this regime. To address this issue, we propose an expansion that substantially improves the accuracy of the analytical expression.
Furthermore, our results motivate us to explore a framework for estimating the fractal dimension of unknown structures. This approach offers an alternative to established methods such as box-counting or sandbox methods. 
Overall, we believe that our analytical work will have broad applicability in systems where every-pair interactions play a crucial role. The insights gained from this study can contribute to a better understanding of various complex systems and facilitate more accurate modelling and analysis in a wide range of disciplines.
\end{abstract}

\maketitle

\newpage

\section{Introduction}

Spatial interactions play an essential role not only in physical systems, like gravitation and electromagnetism, but also in socio-economic systems, where the exchange of ideas drives wealth production  \cite{Schlapfer2014,GlaeserG2009}, and biological systems, where interactions facilitate disease transmission \cite{Brauer2019} and seed dispersion \cite{Trakhtenbrot2014MechanisticMO}, for instance. These interactions typically exhibit a decrease in strength as the spatial distance between the interacting units increases. 
The nature of these interactions can give rise to non-trivial emergent phenomena, including phase transitions and scaling properties \cite{Stanley1987,Tsallisbook}.
In this present work, we aim to investigate two aspects of spatial interaction. 
Firstly, we will focus on how the properties of the medium in which the system is embedded -- specifically, a fractal medium in our study -- either facilitate or hinder the interaction between its components. Secondly, we will examine how the range of interaction, particularly the differentiation between short- and long-ranges of interaction, leads to the emergence of macroscopic effects.

Specifically, short-range interactions are those that act only over relatively small distances. 
Those interactions decay rapidly with distance, meaning that their influence becomes negligible beyond a certain range.
In contrast, long-range interactions are those that can extend over large distances, potentially encompassing the entire system or even infinite ranges. 
The distinction between short- and long-range interactions holds particular importance as it influences the behavior and properties of systems.
Short-range interactions tend to lead to local ordering or clustering of particles, while long-range interactions can give rise to collective behavior and phase transitions. 
However, in many situations, the distinction between short- and long-range interactions is far from obvious. 
This is the case, for instance, in the \emph{Ising model} in \emph{statistical physics} \cite{Kadanoff2000,Yeomans1992}. 
This model considers short-range interactions between spins, meaning that each spin only interacts with its first neighbors. 
However, at the critical temperature, the macroscopic effect is, in fact, a long-range interaction type. This implies that a small perturbation in a single spin can propagate as an avalanche throughout the system with a long-range effect \cite{Kadanoff2000,Tsallisbook}.

A typical quantity of interest in such interactive systems is the average influence that a unit has on the other ones or receives, conversely.
Due to the high level of complexity of arbitrary structures, it can be challenging to derive an analytical expression for these average interactions.
E.g.\ in the context of urban scaling, Dong et al.\ \cite{DongHZL2020} propose a model based on distance-decay interactions between different locations in a city. 
They write that ``\emph{there is no general analytical solution}'' and proceed with numerical simulations.
Considering a structure formed by $N$ components, the computational time to treat all the possible interaction pairs goes as $N^2$, which becomes numerically unfeasible for large $N$.
In this way, an analytical expression which describes the average interaction strength would not only save computational effort but also provide novel insights, e.g.\ about asymptotic behavior. 
And in fact, in the following we present such an analytical expression.

More specifically, we propose an analytical solution to the problem when the structure formed by the spatial distribution of the units can be described by a fractal dimension, and the intensity of the interactions between them depends on the distance following a power-law.
We identify three regimes, (i) for long-range interactions, the average interaction increases as a power-law with the size of the structure; (ii) for short-range interactions, the average interaction increases but saturates when the structure size is sufficiently large; and (iii) between these two regimes, there is one in which the average interaction growth logarithmically with the structure size.
The analytical expression also holds for the case of interactions that increase with the distance, e.g.\ particularly interesting is the average distance between two points of the considered structure.
We restrict our considerations to the case where the cells are spatially arranged forming a fractal structure, and therefore displaying self-similarity.
We have two reasons for this choice. 
First, because we can take analytical advantage of the power-law relation between the number of cells and linear size in a fractal structure. 
And second, because it can represent a good proxy for real-world structures.

We validate the precision of the theoretical expression by analyzing various fractal structures. 
The numerical calculations confirm the three regimes but we observe deviations in the case of short-range interactions.
In order to solve the problem, we propose an expansion refining the theoretical expression and taking discreteness at small scales into account.
Overall, the theoretical expression(s) describe the mean interaction field to a sufficient extent.
The degree of accuracy suggests employing the analytical solution to measure the fractal dimension of unknown structures.
Last, we explore how the expression could be used as part of such a method.
Overall, we include a mathematical discussion of the problem -- readers not interested in such details are referred to the main results Eqs.~\ref{eq_result_anali} and~\ref{eq:final_doublesum}.

\section{Analytical model}
\label{sec:analytical}
Consider a structure consisting of $N$ equal units located in space.
It can, for instance,  be a city formed by $N$ buildings/individuals, or an organism formed by $N$ cells, or a solid formed by $N$ atoms/molecules. 
Suppose these units, hereafter we will simply refer to them as ``cells'', interact with one another in a distance-dependent way.
We denote $I_{ij}$ the \emph{pair interaction intensity} between the cells $i$ and $j$.
It can be the friendship strength between two people, the heat flux between two places, competition/cooperation between two biological cells, etc. 
Here we study distance-dependent interactions following a power-law
\begin{equation}\label{Eq_power_law_decay}
I_{ij} =  \frac{1}{r_{ij}^\gamma} \, ,
\end{equation}
where $r_{ij}$ is the distance between cells $i$ and $j$. 
The exponent $\gamma$ is a parameter of the model. 
When $\gamma>0$ (positive), it represents the decay exponent (gravity model) that governs the range of the interaction. 
The idea here is to simplify all the real-world complexity that governs the cell-cell interaction by a single parameter ($\gamma$). 
However, in our analyses, we will also consider $\gamma<0$ (negative) since this situation also presents some interesting properties, as will be seen in the following sections.

\begin{figure}
	\begin{center}
	a) \includegraphics[width=\columnwidth]{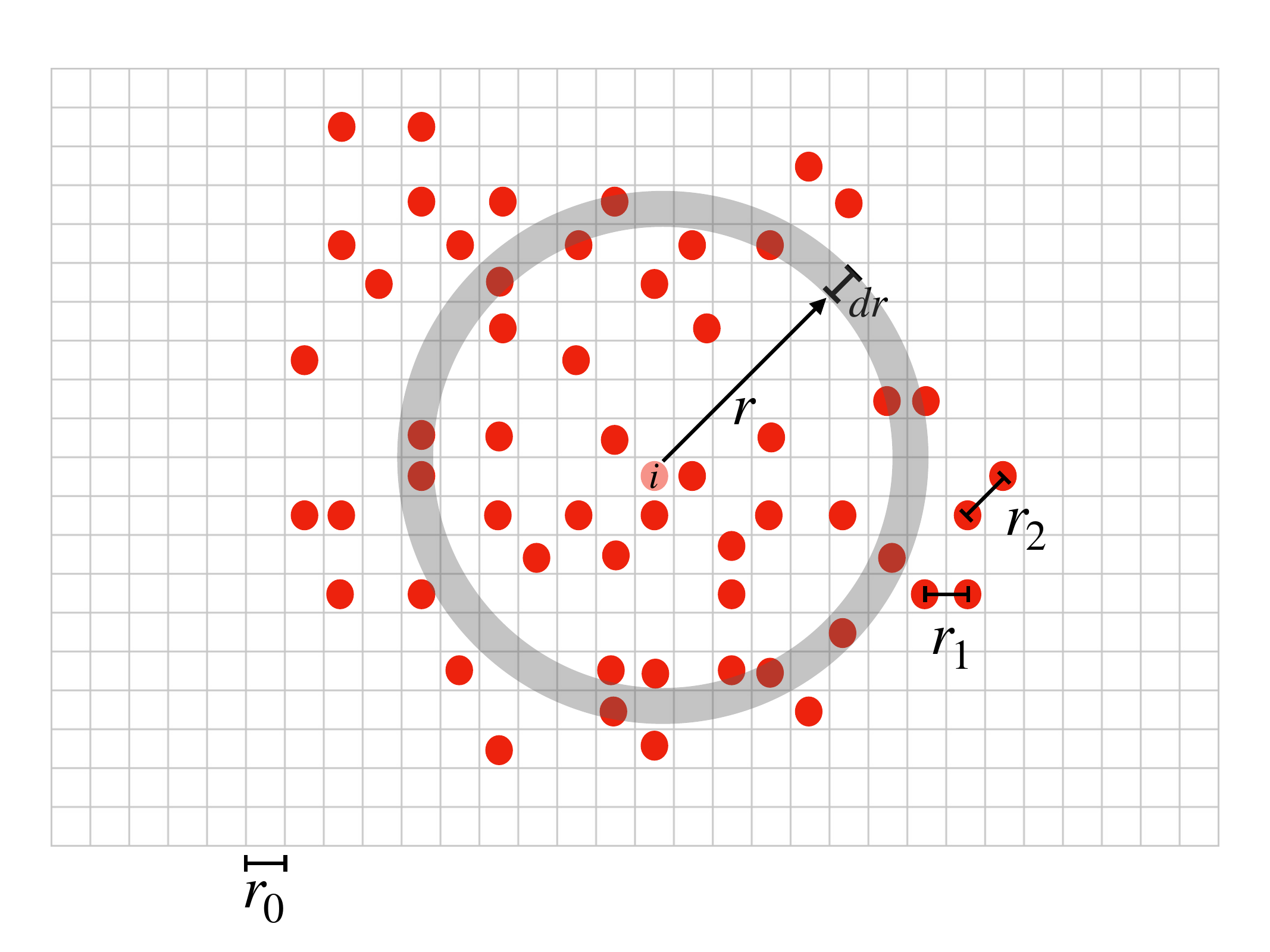}
		b)
  \includegraphics[width=\columnwidth]{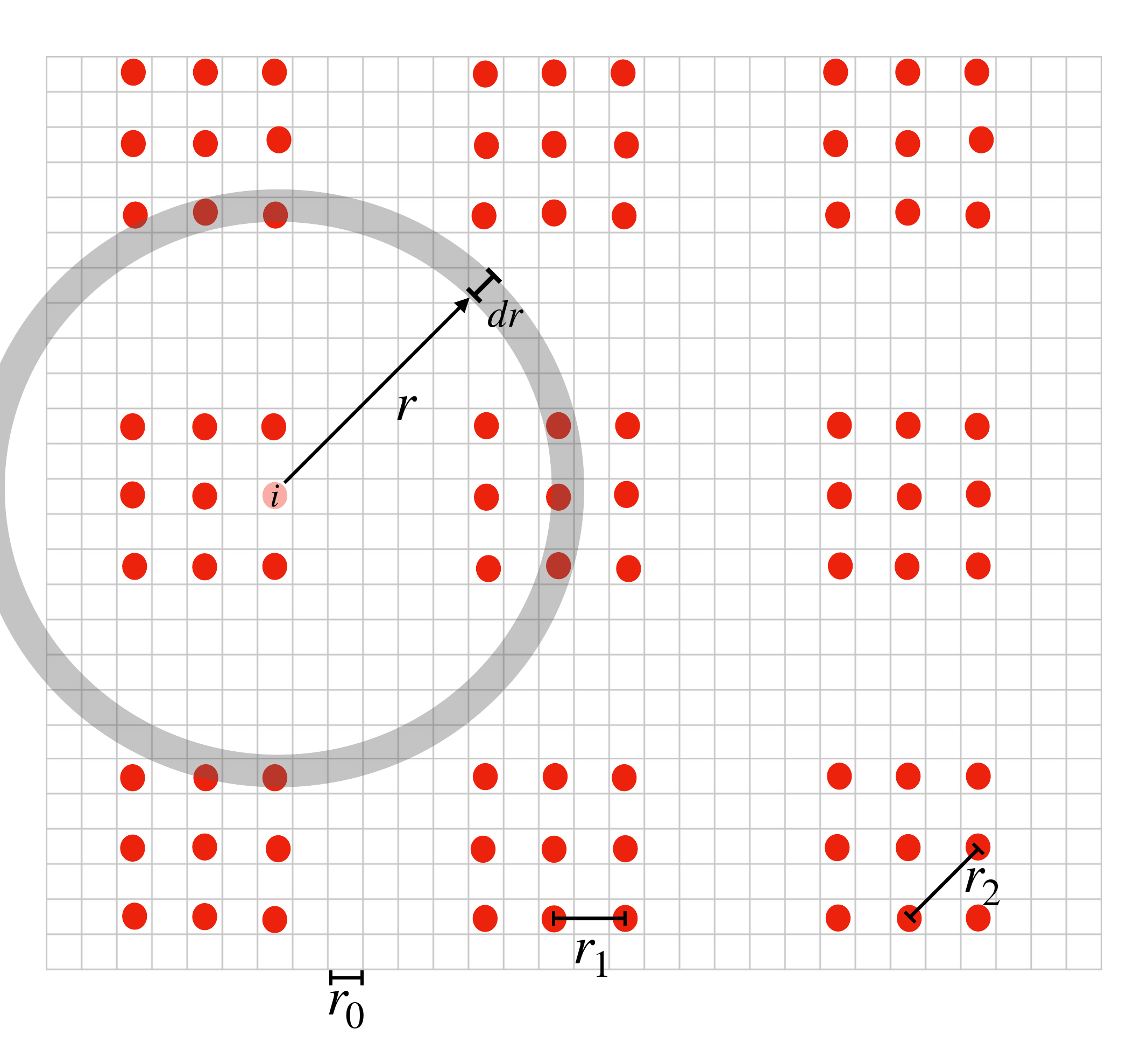}
  \end{center}
	\caption{ \label{fig_cells} 
Examples of two macroscopic structures -- (a) and (b) -- formed by a set of points (red dots) in a grid with a lowest resolution limit $r_0$. 
In a) one has a stochastic structure, where the closest neighbors are  separated by $r_1 = r_0$ and the second closest neighbors are separated by $r_2 = \sqrt{2}r_0$.  
In b) one has a regular structure, where 
 the closest neighbors are  separated by $r_1 = 2 r_0$ and the second closest neighbors are  separated by $r_2 = 2\sqrt{2}r_0$.  That is, the values of $r_1$ and $r_2$ depend on the structure we are analyzing.
While $N(r)$ represents the number of points inside the circle  of radius $r$ (centred in the cell $i$),   $dN$ is the number of points in the ring  (the dashed area) of  radius $r$ and $r + dr$.  
	}	
\end{figure}

The total interaction intensity %
of cell $i$ is then
\begin{equation}
I_i \equiv  \sum_{j\ne i}^N  I_{ij} =  \sum_{j\ne i}^N \frac{1}{r_{ij}^\gamma} \, ,
\label{eq:singlesum}
\end{equation}
where the notation $j\ne i$ indicates that we exclude self-interaction. The quantity 
$I_i$ can be thought of as the \emph{interaction field} acting on the $i$-th cell. 
Moreover, the mean cell interaction intensity $\bar{I_i}$, which we will simply refer to by $I$, can be determined by calculating the sum over the entire structure, i.e.\
\begin{equation}\label{eq:doublesum}
I \equiv \bar{I_i} = \frac{1}{N} \sum_{i=1}^N I_i = \frac{1}{N} \sum_{i=1}^N \sum_{j\ne i}^N \frac{1}{r_{ij}^\gamma} \, .
\end{equation}
It  means that, given any structure composed of spatially arranged units, as the ones presented in Fig.~\ref{fig_cells}, and specific values of $\gamma$, it is possible to compute the double sum described in Eq.~(\ref{eq:doublesum}) and, consequently, to obtain the mean cell interaction intensity $I$.

To treat the problem analytically, we assume that the sum in Eq.~(\ref{eq:singlesum}) can be written as an integral (continuum approximation)
\begin{equation}\label{eq_aproxx}
I_i = \sum_{j\ne i}^N \frac{1}{r_{ij}^\gamma} \quad \to  \quad  I_i^{\rm{theo}} \equiv \int  \frac{1}{r^\gamma} dN(r)
\, .
\end{equation}
Note that $I_i$ represents the numerical value, while $I_i^{\rm{theo}}$ is the theoretical estimation of this quantity.
Moreover, $dN(r)$ is the number of cells that are at a distance between $r$ and $r + dr$ from the cell $i$. In Fig.~\ref{fig_cells}, $dN(r)$ includes all the cells in the gray area.

Next, we take advantage of the considered structure being a fractal.
Then one can say that the total number of cells inside a circle with radius $r$, denoted $N(r)$, obeys
\begin{equation}\label{eq_fractal}
N(r) = N_0 r^{D_f} \, ,
\end{equation}
where $D_f$ is the fractal dimension of the object \cite{BundeHFractalsScience1994-1} and $N_0$ is a constant (idealized, the average number of cells inside a circle of radius $r = 1$). 
If $R_{\rm{max}}$ is the radius of a circle that covers the entire object, 
then $N(r= R_{\rm{max}})$ is the total number of cells $N$. 
Consequently, one can write $N$ as a function of the parameters of the model, that is'
\begin{equation}\label{eq:Rmax}
N = N(r= R_{\rm{max}}) = N_0 R_{\rm{max}}^{D_f} \, .
\end{equation}
Of course, Eq.~(\ref{eq:Rmax}) represents an idealization, since (due to boundary effects) 
it is not possible to simultaneously satisfy the fractal relation Eq.~(\ref{eq_fractal}) and to have a radius ($r = R_{\text{max}}$) that covers the entire structure.
However, for a sufficiently large structure boundary effects can be minimized and the error of Eq.~(\ref{eq:Rmax}) becomes negligible.

From the derivative of Eq.~(\ref{eq_fractal}) we obtain $dN (r)$, the number of cells between $r$ and $r + dr$, which is 
\begin{equation}\label{eq:dNDf}
dN (r) = N_0 D_f r^{D_f -1} dr
\, .
\end{equation}
Inserting Eq.~(\ref{eq:dNDf}) in the integral of Eq.~(\ref{eq_aproxx}) we get
\begin{equation}
I_i^{\rm{theo}} = N_0 D_f \int_{r_1}^{R_{\rm{max}}} r^{D_f - \gamma -1} dr
\, , 
\end{equation}
where $r_1$ is the distance between the first neighbors,
resulting in
\begin{equation}
I_i^{\rm{theo}} = \frac{N_0 D_f}{(D_f - \gamma)} \left[ R_{\rm{max}}^{D_f - \gamma} - r_1^{D_f - \gamma}\right]
\, .
\end{equation}
Finally, we can write this expression in terms of the size (total number of cells) $N$.
Using Eq.~(\ref{eq:Rmax}) we obtain $R_{\rm{max}} = {(N/N_0)}^{\frac{1}{D_f}}$ and then 
\begin{equation}\label{eq_result_anali}
I^{\rm{theo}} \simeq I_i^{\rm{theo}} = 
\frac{N_0 }{(1 - \frac{\gamma}{D_f})} \left[ \left( \frac{N}{N_0} \right)^{1- \frac{\gamma}{D_f}} -  r_1^{D_f - \gamma} \right]
\, .
\end{equation}
This result represents the theoretical expression for the interaction field in cell $i$, which, in turn, is a function of $\gamma$ and $N$, given the parameters $D_f$, $N_0$, and $r_1$. 
In compact notation, we can express this result as $I_i^{\rm{theo}} = I_i^{\rm{theo}}(N, \gamma|D_f, N_0, r_1)$. 
The theoretical interaction field in cell $i$ depends solely on global (macroscopic) information, indicating that it is identical for all other cells. 
This observation highlights that the result obtained from Eq.~(\ref{eq_result_anali}) corresponds to a type of \emph{mean-field approximation}.
For the sake of simplicity, moving forward, we will omit the index~$i$ and refer to the interaction field as $I^{\rm{theo}}$. 
A similar calculus in the context of cellular growth was done in \cite{mombach2002,DOnofrio2009a}, and for a one-dimensional simpler version of this analysis, see \cite{ribeiro1d}.

\begin{figure*}
	\begin{center}
	\includegraphics[width=\textwidth]{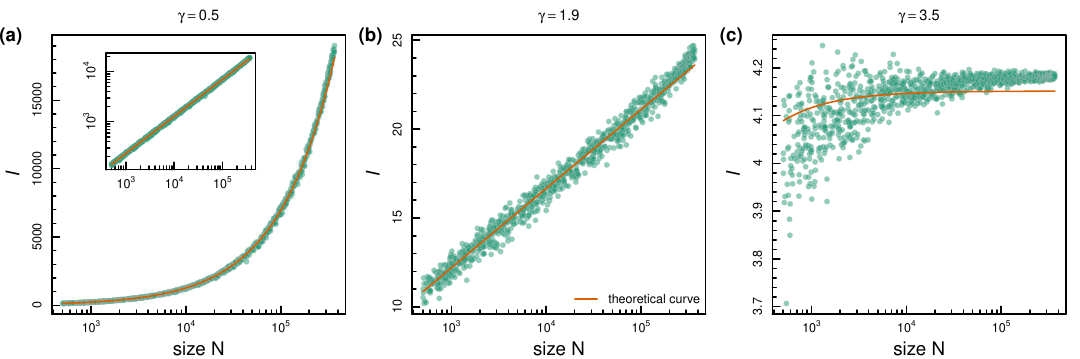}
    \caption{\label{fig:I_N_plot}
Regimes of size dependence of the mean interaction field. 
The mean interaction field $I$ is plotted as a function of the size $N$ of  structures generated by realizations of percolation clusters using the Leath algorithm, which has fractal dimension $D_f= \frac{91}{48}\approx 1.896$.
In all panels,
the dots represent numerical values obtained performing Eq.~(\ref{eq:doublesum}), and the solid line represents the theoretical curve according to Eq.~(\ref{eq_result_anali}) using $r_1 = 1$, where $N_0$ is estimated by non-linearly fitting $I$ as a function of $N$ with $\gamma$ and $D_f$ known.
All panels are on a semi-logarithmic scale, but the inset on panel~(a) is on a log-log scale.
(a) For $\gamma=0.5 < D_f$, we are treating long-range interactions implying a power-law relation Eq.~(\ref{Eq_long-range}),
as evidenced by the straight line in the inset (log-log scale).
(b) For $\gamma=1.9 \approx D_f$, we are treating the transition between short- and long-range interactions, i.e.\ a logarithmic relation Eq.~(\ref{Eq_gammaeqDf}), 
recognizable as a straight line in semi-logarithmic scale.
(c) For $\gamma=3.5 > D_f$ we are treating short-range interactions implying a saturation Eq.~(\ref{Eq_short-range}) with disagree between theoretical and numeric values. This discordance 
is discussed in Sec.~\ref{sec_num_val}.
}
\end{center}
\end{figure*}

\subsection{Exploring the analytical result}
\label{ssec:exploring}
We can extract some insights about how the system behaves by exploring the solution Eq.~(\ref{eq_result_anali}).
The first thing to be noted is that depending on the values of $\gamma$ and $D_f$, for asymptotic large $N$, we can distinguish two regimes separated by the special case $\gamma = D_f$. 
These three cases are discussed in the following and are illustrated in Fig.~\ref{fig:I_N_plot}.

\begin{itemize}

\item For $\gamma<D_f$ the exponent in Eq.~(\ref{eq_result_anali}) is $1-\frac{\gamma}{D_f}>0$, so that for $N\gg N_0$ we obtain 
\begin{equation}\label{Eq_long-range}
I^{\rm{theo}} \sim N^{1-\frac{\gamma}{D_f}} \, ,   
\end{equation}
i.e.\ a power-law dependence of the interaction field with the size, as illustrated in Fig.~\ref{fig:I_N_plot}(a).
As the interactions are noticeable on the entire structure, this situation ($\gamma<D_f$) can be called \emph{long-range interaction regime}.
In these $\gamma$ values, to determine the mean interaction field theoretically, it is sufficient to know global (macroscopic) information, i.e.\ $D_f$ and $N$; any local (microscopic) information of the structure is irrelevant to determine $I$.  
An approach of this kind, described by Eq.~(\ref{Eq_long-range}), was used to explain the origin of urban scaling in socio-economic variables \cite{ribeirocity2017,Ribeiro2021b}.

\item For $\gamma>D_f$ the exponent in Eq.~(\ref{eq_result_anali}) becomes $1-\frac{\gamma}{D_f}<0$ so that for $N\gg N_0$ we obtain 
\begin{equation}\label{Eq_short-range}
I^{\rm{theo}} \to \frac{N_0 D_f}{(\gamma-D_f)}r_1^{D_f-\gamma} \, ,   
\end{equation}
i.e.\ an asymptotic value.
This is the case of \emph{short-range interaction regime}, i.e.\ there is a typical range of interactions.
In fact, these $\gamma$ values represent a situation where the details at the local (microscopic) level play an essential role in the mean interaction field. 
In Fig.~\ref{fig:I_N_plot}(c) the saturation of $I$ for large $N$  appears in both  theoretical and  numerical estimates.
Reasons for deviations are discussed in Sec.~\ref{sec_num_val}.

\item In between, for $\gamma=D_f$, the asymptotic relation is logarithmic 
\begin{equation}\label{Eq_gammaeqDf}
I^{\rm{theo}} \sim N_0\ln\left(\frac{N}{N_0}\right)\sim \ln N \, ,
\end{equation}
as can be seen by the straight line in the semi-log plot Fig.~\ref{fig:I_N_plot}(b).

\end{itemize}

We can also consider asymptotic values of $\gamma$. 
This means we want to verify how $I^{\rm{theo}}$ behaves with $\gamma$ when extreme values of this decay exponent are analyzed, keeping $N$ fixed and large. 
\begin{itemize}
\item  In the limit of $\gamma \rightarrow -\infty$ one can show that the mean interaction field increases exponentially with $\gamma$ as 
\begin{equation}
\label{eq:slopegammainfty}
I^{\rm{theo}}(\gamma)~\sim~e^{-\frac{\gamma}{D_f} \ln(N/N_0)} \, .
\end{equation}
Accordingly, the slope is $\frac{d\ln{I}}{d\gamma}=-\frac{1}{D_f}\ln{\frac{N}{N_0}}$.

\item In the limit of $\gamma \rightarrow \infty$ there are two cases for the mean interaction field. 
(i) For $r_1 \ne 1$ it decreases exponentially with $\gamma$ following 
\begin{equation}I^{\rm{theo}}(\gamma)~\sim~e^{- \gamma\ln(r_1)} \, .
\end{equation} 
This means it vanishes for sufficiently large $\gamma$. 
The slope is $\frac{d\ln{I}}{d\gamma} = -\ln{r_1}$.
(ii) For $r_1=1$ the mean interaction field decreases with $\gamma$ following $I^{\rm{theo}}(\gamma)~\sim~\frac{N_0 D_f}{\gamma}$.
Then the slope is $\frac{d\ln{I}}{d\gamma} = -1/\gamma$.

\end{itemize}

Last we want to mention two special cases.
(i) For $\gamma = 0$ the physical distance does not matter and Eq.~(\ref{eq_result_anali}) yields 
\begin{equation}\label{eq:gamma0}
    I^{\rm{theo}} = N - r_1^{D_f} \, .
\end{equation}
Moreover, $r_1 =1$ implies $I^{\rm{theo}} = N - 1$, which is expected when each cell interacts with every other cell except with itself. 
(ii) For $\gamma = -1$, the interaction field is directly related to the average distance  $\langle r \rangle$ between cells by the form $I^{\rm{theo}} = N \langle r \rangle$.
From Eq.~(\ref{eq_result_anali}) we obtain the average distance between cells of a fractal structure, given by
\begin{equation}
\langle r \rangle  = 
\frac{N_0}{N(1 + \frac{1}{D_f})} \left[ \left( \frac{N}{N_0} \right)^{1 + \frac{1}{D_f}} -  r_1^{D_f +1} \right]
\, .
\end{equation}
For $ N \gg N_0 $ this leads to  
\begin{equation}
  \langle r \rangle \sim N^{\frac{1}{D_f}} \, ,
\end{equation}
justifying some consideration usually done in the literature \cite[e.g.]{PrietoCurielPA2023,BettencourtLMA2013}.

For the sake of completeness, in Appendix~\ref{app:exploring} we discuss further properties of Eq.~(\ref{eq_result_anali}), and in Appendix~\ref{sec:genlog} we write this solution in terms of the generalized logarithm.

\subsection{Collapse}
\label{ssec:collapse}

In the following, we want to reinspect Eq.~(\ref{eq_result_anali}) and propose a ``collapse'' (we use quotation marks here since below we see in which situation the collapse fails).
For $N\gg~N_0$ and $\gamma < D_f$ the first term of Eq.~(\ref{eq_result_anali}) dominates, and then it is possible to write
$I^{\rm{theo}}~\approx~\frac{N_0 D_f}{(1 - \frac{\gamma}{D_f})} \left( \frac{N}{N_0} \right)^{1- \frac{\gamma}{D_f}}$, and consequently
\begin{equation}\label{eq:collpase}
I^{\rm{theo}} \approx \frac{N_0}{q}\left( \frac{N}{N_0} \right)^q
\quad \text{with} \quad q=1-\frac{\gamma}{D_f} \, .
\end{equation}
This means, if we keep $N$ and $N_0$ fixed, then  
$I^{\rm{theo}} = I^{\rm{theo}}(q)$ when  $q>0$ ($\gamma < D_f$). More specifically, if we plot $I^{\rm{theo}}$ on the vertical axis and $q$ on the horizontal one, as presented in Fig.~(\ref{fig:collapse_test}-b), then the values of different structures should fall on the same \emph{collapsed curve} (i.e.\ with different values of $D_f$, but keeping $N$ and $N_0$ fixed).
More details in section~(\ref{sec_numerical_val}).

A conclusion that we can draw is that, 
in the long-range regime ($\gamma < D_f$, $q>0$), 
$q$ is the natural variable of the interaction field, as also suggested in Appendix~\ref{sec:genlog}. %
But we continue using $\gamma$ in the remainder of the paper because we are also interested in the short-range regime ($\gamma > D_f$, $q<0$).

\subsection{Relation to the Grassberger-Procaccia algorithm}

The mathematical structure of Eq.~(\ref{eq:doublesum}) suggests similarities between our approach  and  the  Grassberger-Procaccia algorithm \cite{GrassbergerP1983prl,GrassbergerP1983physd}. 
This algorithm is based on the quantity
\begin{equation}
\label{eq:GPdef}
C (a) = \frac{2}{N(N-1)} \sum_{i=1}^N \sum_{j\ne i} \Theta{(a-r_{ij})} \, ,
\end{equation}
which is calculated considering $N$ cells spatially arranged 
(consistent with the work in hand),
where $\Theta(\cdot)$ is the \emph{Heaviside step function} and $a$ is a parameter.
In the original work \cite{GrassbergerP1983prl},  the authors show that, in the case of a fractal structure, the power-law relation 
\begin{equation}
\label{eq:GPfract}
C(a) \sim a^{D_c} 
\end{equation}
is expected. Here $D_c$ is the so-called \emph{correlation fractal dimension}. 
In Appendix~\ref{sec:detailsGP}, we show that, in the case of a mono-fractal, the result expressed by Eq.~(\ref{eq:GPfract}) can also be derived from our framework.

If we compare Eq.~(\ref{eq:GPdef}) with Eq.~(\ref{eq:doublesum}), then we notice that both consist of a double-sum, and they only differ in their argument (apart from the pre-factors).
That is, in the case of the Grassberger-Procaccia algorithm
the pair interaction intensity $I_{ij}$ is given by 
$I_{ij} = \Theta{(a-r_{ij})}$, and in our framework it is $I_{ij} = \frac{1}{r_{ij}^\gamma}$, Eq.~(\ref{Eq_power_law_decay}).
We can also observe that both have a free parameter, i.e.\ $a$ in the Grassberger-Procaccia algorithm and $\gamma$ in our approach.
This motivates us to ask under which conditions both approaches are equivalent.
That is the case, when both $I^{\textrm{theo}}$ are the same.
In Appendix~\ref{sec:detailsGP}, we show that this is given when
\begin{equation}\label{eq:GPagamma}
    \gamma =   D_f - D_f^2\frac{\ln a}{\ln N} + {\rm cte} \cdot \frac{D_f}{\ln N} \, ,
\end{equation}
if we consider asymptotic behavior of Eq.~(\ref{eq:GPfract}), the long-range case in our framework Eq.~(\ref{Eq_long-range}), and keeping $N$ fixed.
This means if Eq.~(\ref{eq:GPagamma}) holds true, then the long-range case of our approach is equivalent to the Grassberger-Procaccia algorithm.
Our $\gamma$ essentially follows a (negative) logarithm relationship with $a$, whereas $a\rightarrow \infty$ implies $\gamma \rightarrow -\infty$, i.e.\ the long-range regime (self-consistent).

\section{Numerical validation and refinement}
\label{sec_numerical_val}

Next we want to validate our theoretical expression.
Therefore, we generate fractal structures, calculate the interaction intensity $I$ numerically, and compare the outcome with the theoretical value according to Eq.~(\ref{eq_result_anali}).

\subsection{Fractal structures}
\label{ssec:fractals}
We consider regular and random fractals for the analysis.
First, the regular fractals are generated iteratively, literally imposing self-similarity.
Second, as random fractal, we study percolation clusters.

Specifically, we create the regular fractal structures from a base pattern on a $5 \times 5$ square grid (all base patterns are displayed in Fig.~\ref{fig:base_structure}).
As illustrated in Fig.~\ref{fig:regular_fractal_growth}, starting from the base pattern, we grow the fractal structure by iteratively replacing each occupied cell with the base structure itself. 
If the number of cells of the base structure is $A$, then the pattern at the $k^\text{th}$ iteration consist of $A^{k+1}$ cells, extending on the grid with a length of $5^{k+1}$.
In this way, the patterns grow with a well-defined fractal dimension, which can be calculated as $D_f = - \frac{\ln A^{k+1} - \ln 1}{\ln 1 - \ln 5^{k+1}} = \frac{\ln A}{\ln 5}$, and which depends only on the number of cells of the base pattern.

We grow each fractal structure with up to 4 iterations and store structures at the 3$^\text{rd}$ and 4$^\text{th}$ iteration for the analyses. 
Due to excessive size beyond the 2$^\text{nd}$ iteration, Fig.~\ref{fig:regular_fractal_growth} only shows the base pattern, and fractal structures at the 1$^\text{st}$ and 2$^\text{nd}$ iterations (the actual structures being analyzed are not shown).

In addition, we consider the Vicsek-fractal, which is created from a `+' in a $3 \times 3$ base pattern. %
Growing from the base pattern by 4 steps leads to a structure with a size of 3,125, and the structure has a fractal dimension of $\frac{\ln 5}{\ln 3} \approx 1.465$.

As random fractals we generate percolation clusters via the Leath algorithm \cite{LeathPL1976,BundeHFractalsDisorderedSystems1991-2}.
Specifically, we run the algorithm on a square grid of size $1,000\times 1,000$, with the occupation probability $p = p_c = 0.592746$. 
This percolation cluster has a theoretical fractal dimension of $D_f^\text{perc}=\frac{91}{48} \approx 1.8958$ 
\cite{BundeHFractalsDisorderedSystems1991-2}. 
We generate 30,000 structures with a size larger than 500. 
To evenly cover the range of sizes, we sample 1,000 structures from the 30,000 realizations, the sizes of these 1,000 examples range from 504 to 369,207.

\subsection{Numerical validation}
\label{sec_num_val}
After generating the fractals, we numerically calculate the interaction intensity~$I$ as defined in Eq.~(\ref{eq:doublesum}).
The computation of $I$ requires a choice of $\gamma$, and we sample 151 values of it 
in the interval $\gamma \in [-5,10]$.

\begin{figure*}
	\begin{center}
	\includegraphics[width=0.8\textwidth]{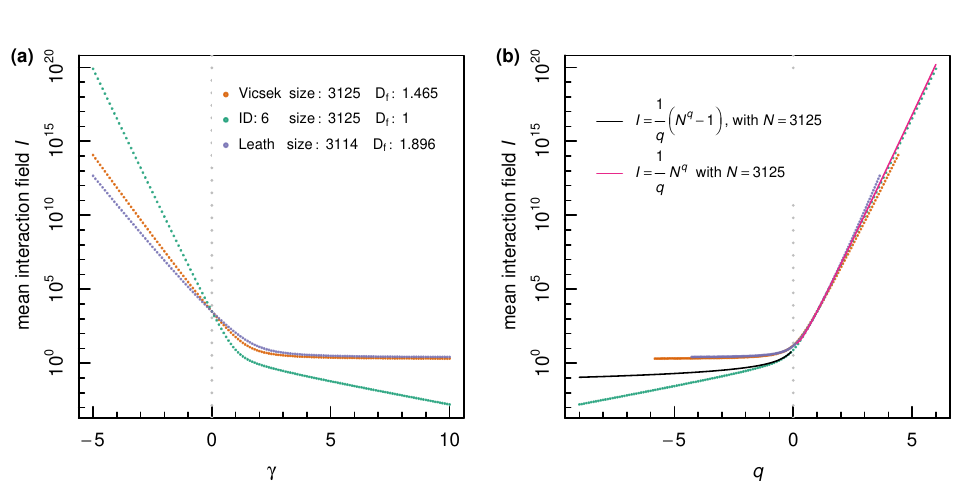}
    \caption{
    \label{fig:collapse_test}
Comparison of the mean interaction field of different structures.
 The numerically calculated mean interaction field $I$ values (from Eq.~(\ref{eq:doublesum}))
 are plotted against 
 (a)  $\gamma$ and 
(b) $q=1-\frac{\gamma}{D_f}$. %
The curves are shown for three different structures with similar size $N$ but different fractal dimensions.
Orange curves are from the Vicsek-fractal, green curves are from the structure~\#6 as in Fig.~\ref{fig:base_structure}, and purple curves are from a percolation cluster generated with the Leath algorithm.
In panel~(b), the pink line represents the curve $I = \frac{1}{q}N^q$, i.e.\ Eq.~(\ref{eq:collpase}) with $N_0 = 1$, $N=3125$. %
Since $I$ becomes negative for $q<0$ in Eq.~(\ref{eq:collpase}), we use $I = \frac{1}{q}(N^q -1)$ for $q<0$ (black line), i.e.\ Eq.~(\ref{eq_result_anali}) with $N_0 = 1$, $N=3125$, and $r_1 = 1$).
For $q>0$ ($\gamma < D_f$, long-range regime), we find that the structures, besides having different fractal dimensions, exhibit approximately the same behaviour following the theoretical expression. 
It confirms that $q$ is the natural variable in the long-range interaction regime.
Minor deviations of the curves are certainly due to $N_0$ that influences the slope for $\gamma \rightarrow -\infty$, see Eq.~(\ref{eq:slopegammainfty}).
For $q<0$ ($\gamma > D_f$, short-range regime), however, none of the cases agrees with the theoretical prediction, and the structures exhibit different behaviour. 
}
\end{center}
\end{figure*}

In Fig.~\ref{fig:collapse_test}(a) we plot the values of $I$ against $\gamma$.
For positive and negative $\gamma$ the curves diverge from each other -- they cross in $\gamma=0$ since this case basically corresponds to counting the number of occupied cells, see Eq.~(\ref{eq:gamma0}).

To align the curves, we can employ the collapse as proposed in Sec.~\ref{ssec:collapse}.
Thus, in Fig.~\ref{fig:collapse_test}(b), we plot the values of $I$ against $q=1-\frac{\gamma}{D_f}$ for three structures with approximately the same size but different fractal dimensions.
For $q>0$ ($\gamma < D_f$, long-range regime), we find that the structures exhibit the same behaviour approximately following the theoretical expression Eq.~(\ref{eq_result_anali}), and consequently Eq.~(\ref{eq:collpase}) with an arbitrary choice $N_0=1$.
The collapse of the curves on the right side of Fig.~\ref{fig:collapse_test}(b) confirms that $q$ is the natural variable for a system in a long-range interaction regime (still, in the remaining figures, we use $\gamma$ as we find it more intuitive).
Nevertheless we can observe that the curves on the right side of Fig.~\ref{fig:collapse_test}(b) exhibit minor deviations, which are certainly due to $N_0$ that influences the slope when $\gamma \rightarrow -\infty$, see Eq.~(\ref{eq:slopegammainfty}).

However, for $q<0$ ($\gamma > D_f$, short-range regime), we see clear deviations among the three examples.
None of the cases agrees with the theoretical curve.
Moreover, the curves exhibit different asymptotic slopes, and while the Vicsek-fractal and the  percolation cluster (Leath algorithm) exhibit a horizontal asymptote, the fractal structure \#6 exhibits an inclined one.
Accordingly, Eq.~(\ref{eq_result_anali}) fails in the short-range regime.

To understand why Eq.~(\ref{eq_result_anali}) fails in the short-range regime ($\gamma > D_f$), we can first ask, why do we obtain different asymptotic slopes.
Apparently, for $\gamma \rightarrow \infty$ only cells at distance $d=1$ (Von~Neumann neighborhood) from a specific cell $i$ make a contribution and all others are omitted.
If the cells of the considered fractal structure have none of the four immediate neighbors, e.g.\ a Cantor-like fractal (like structure \#27 and \#28 in Fig.~\ref{fig:base_structure}), then $I$ goes to zero for asymptotic large $\gamma$.
If the cells of the considered fractal structure do have one or more immediate neighbors, then $I$ corresponds to the average number of nearest neighbors for asymptotic large $\gamma$.
From Sec.~\ref{ssec:exploring} we know that the asymptotic slope for $\gamma\rightarrow\infty$ is $\frac{d\ln{I}}{d\gamma} = -\ln{r_1}$.
If $r_1=1$, i.e.\ there are immediate neighbors, then we obtain a horizontal asymptote (slope $\sim 0$).
If $r_1>1$, i.e.\ there are no immediate neighbors, then we obtain an inclined asymptote.
Thus, the failure of Eq.~(\ref{eq_result_anali}) in the short-range regime ($\gamma > D_f$) is an artefact due to the discreteness of the underlying grid, i.e.\ the assumption Eq.~(\ref{eq_aproxx}) is not valid.

\subsection{Refinement}
The disagreement between numerical and theoretical prediction in the short-range regime motivates us to refine Eq.~(\ref{eq_result_anali}), taking the local characteristics into account.
We begin by separating the sum in
Eq.~(\ref{eq:singlesum}) into two other sums, as 
\begin{equation}
I_i = \sum_{j \in N_1^i} \frac{1}{r_1^\gamma} + \sum_{j \notin N_1^i} \frac{1}{r_{ij}^\gamma}
\, ,
\end{equation}
where $N_1^i$ is the number of nearest neighbors of cell $i$, that is the ones that are at distance $r_1$ from $i$, and employing the integral from Eq.~(\ref{eq_aproxx}) yields 
\begin{equation}\label{eq:sumint}
I_i^{\rm{theo}} = \sum_{j \in N_1^i} \frac{1}{r_1^\gamma} + \int_{r = r_2}^{R_{\rm{max}}} \frac{1}{r^\gamma} dN(r)
\, ,
\end{equation}
where $r_2$ is the distance to the second nearest neighbours (see representation in Fig.~(\ref{fig_cells}).

Considering the short-range interaction case ($\gamma>D_f$), and that for $\gamma \gg D_f$ only the nearest neighbors have a contribution, leads to
\begin{equation}
I_i \approx \sum_{j \in N_1^i} \frac{1}{r_1^\gamma} \approx N_1^i \frac{1}{r_1^\gamma}
\, .
\end{equation}
Replacing the sum in Eq.~(\ref{eq:sumint}) and integrating as before, leads to
\begin{equation}
I_i^{\rm{theo}} = \frac{N_1^i}{r_1^\gamma} + \frac{N_0 D_f}{(D_f - \gamma)} \left[ \left( \frac{N}{N_0} \right)^{1- \frac{\gamma}{D_f}} -  r_2^{D_f - \gamma} \right]
\, .
\label{eq:r1r2}
\end{equation}
Assuming we have a sufficiently large structure, we obtain
\begin{equation}
\begin{aligned}
    I^{\rm{theo}} & = \frac{1}{N}\sum_{i =1}^N I_i^{\rm{theo}}\\
      & \approx \frac{N_1}{r_1^\gamma} + \frac{N_0 D_f}{(D_f - \gamma)} \left[ \left( \frac{N}{N_0} \right)^{1- \frac{\gamma}{D_f}} -  r_2^{D_f - \gamma} \right]
    \, ,
    \label{eq:final_doublesum}
\end{aligned}
\end{equation}
where $N_1$ is the average number of nearest neighbours at a distance $r_1$ that each grid has.

Additionally, we can take advantage of the case $\gamma=0$ (see around  Eq.~(\ref{eq:gamma0})).
On the one hand, Eq.~(\ref{eq:final_doublesum}) implies $I^{\rm{theo}} = N_1 + N - N_0 r_2^{D_f}$ when $\gamma=0$.
On the other hand, $\gamma=0$ implies $I=N-1$ in Eq.~(\ref{eq:doublesum}).
Thus, together we have
\begin{equation}
\label{eq:N0N1}
    N_0 =\frac{N_1 + 1}{r_2^{D_f}}
    \, ,
\end{equation}
which can be used in Eq.~(\ref{eq:final_doublesum}).

\begin{figure*}
	\begin{center}
	\includegraphics[width=0.8\textwidth]{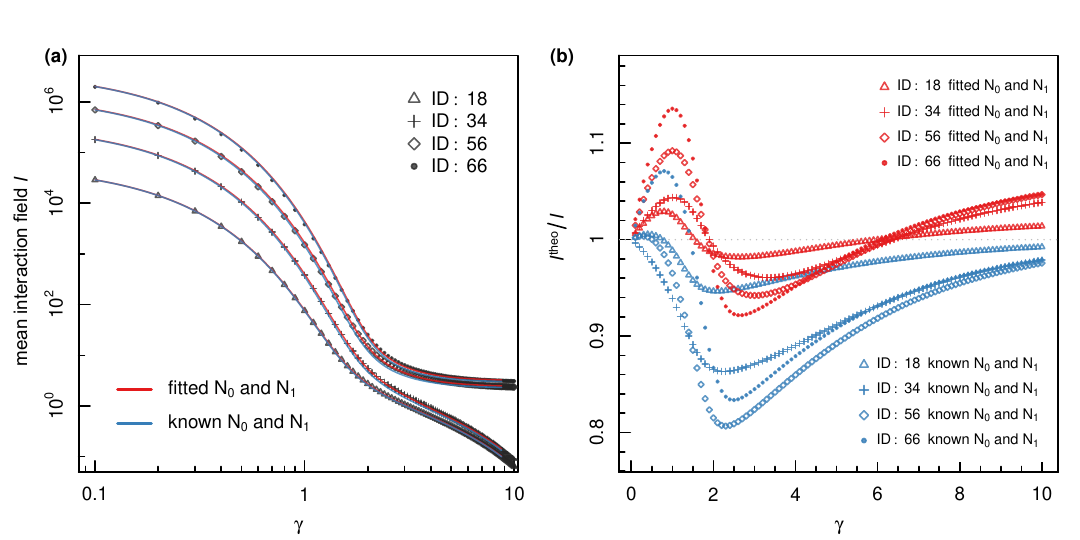}
    \caption{
Comparison of numerical and refined theoretical values of the mean interaction field for different structures.
(a) The mean interaction field is plotted as a function of $\gamma$ in a double logarithmic scale.
Symbols represent the numerical values for four different structures as indicated in the legend (see Fig.~\ref{fig:base_structure}).
The structures differ in size ($N=9^5$, $13^5$, $17^5$, $21^5$, respectively), in fractal dimension ($D_f =$ 1.37, 1.59, 1.76, 1.89, respectively), as well as $r_1$ ($\sqrt{2}, \sqrt{2}, 1, 1$) and $r_2$ ($2, 2, \sqrt{2}, \sqrt{2}$) values.
Solid lines represent curves according to Eq.~(\ref{eq:final_doublesum}).
Red and blue curves differ in the way that $N_0$ and $N_1$ are obtained.
For the red curves, they are treated as fitting parameters.
For the blue curves, $N_1$ is calculated numerically via its definition as the average number of nearest neighbours at a distance $r_1$, and $N_0$ is calculated via Eq.~(\ref{eq:N0N1}).
(b) To better visualize the agreement of numerical and theoretical curves, we plot the quotient of theoretical and numerical values of the mean interaction field  as a function of $\gamma$.
Symbols and colours are the same as in panel~(a).
    }
    \label{fig:regular_fractal_fitting}
\end{center}
\end{figure*}

Next, we want to validate Eq.~(\ref{eq:final_doublesum}) numerically, i.e.\ compare this refined theoretical expression with the respective numerical values.
Therefore, we consider two approaches to applying Eq.~(\ref{eq:final_doublesum}), which differ in the way how we deal with unknown parameters.
The size $N$ is known, and it is given by the number of occupied cells; the values of $r_1$ and $r_2$ are known (can be inferred from the structures); and the fractal dimension is also known by its theoretical values, Sec.~\ref{ssec:fractals}. 
It remains to find the values of $N_0$ and $N_1$.
In the first approach, we apply non-linear curve fitting and treat $N_0$ and $N_1$ as free parameters.
In the second approach, we calculate $N_1$ numerically, following its definition as the average number of nearest neighbours at a distance $r_1$, and calculate $N_0$ via Eq.~(\ref{eq:N0N1}).

In Fig.~(\ref{fig:regular_fractal_fitting}), we consider four fractal structures and apply Eq.~(\ref{eq:final_doublesum}) following these two approaches.
Figure~\ref{fig:regular_fractal_fitting}(a) depicts the numerical values $I$ together with the theoretical $I^{\rm{theo}}$ as a function of $\gamma$. 
Visually, barely any difference can be seen, which suggests that the refined Eq.~(\ref{eq:final_doublesum}) also captures the short-range regime $\gamma>D_f$.
Since the vertical axis in Fig.~\ref{fig:regular_fractal_fitting}(a) is logarithmic, it is difficult to assess deviations.
Therefore, in Fig.~\ref{fig:regular_fractal_fitting}(b) we plot the ratio of theoretical and numerical $I$.
As can be seen, when treating $N_0$ and $N_1$ as fitting parameters, the theoretical values show deviations of approximately 10\,\% in both directions.
When we pre-calculate $N_0$ and $N_1$, we find deviations up to 20\,\% but mostly in the same direction (theoretical value being too small).
However, in this latter case, the asymptotic values agree, which suggests that Eq.~(\ref{eq:final_doublesum}) works but spurious discreteness effects still play a role.

\section{Estimating the fractal dimension}

Last we want to point out that Eq.~(\ref{eq:final_doublesum}) can also be used to estimate the fractal dimension of a structure.
For our \emph{proof of concept} we consider percolation clusters as obtained from the Leath algorithm (Sec.~\ref{ssec:fractals}).
Therefore, we numerically calculate $I$ and fit Eq.~(\ref{eq:final_doublesum}) with known $N$, $N_0$, $N_1$, $r_1$, and $r_2$.
That is, employing non-linear curve fitting, the only parameter to be tuned is the fractal dimension $D_f$, which represents our estimate.
In addition, we exclude the $\gamma<0$ from the analyses
(in log-scale, it would not work) 
without loss of generality.

\begin{figure*}
	\begin{center}
	\includegraphics[width=0.8\textwidth]{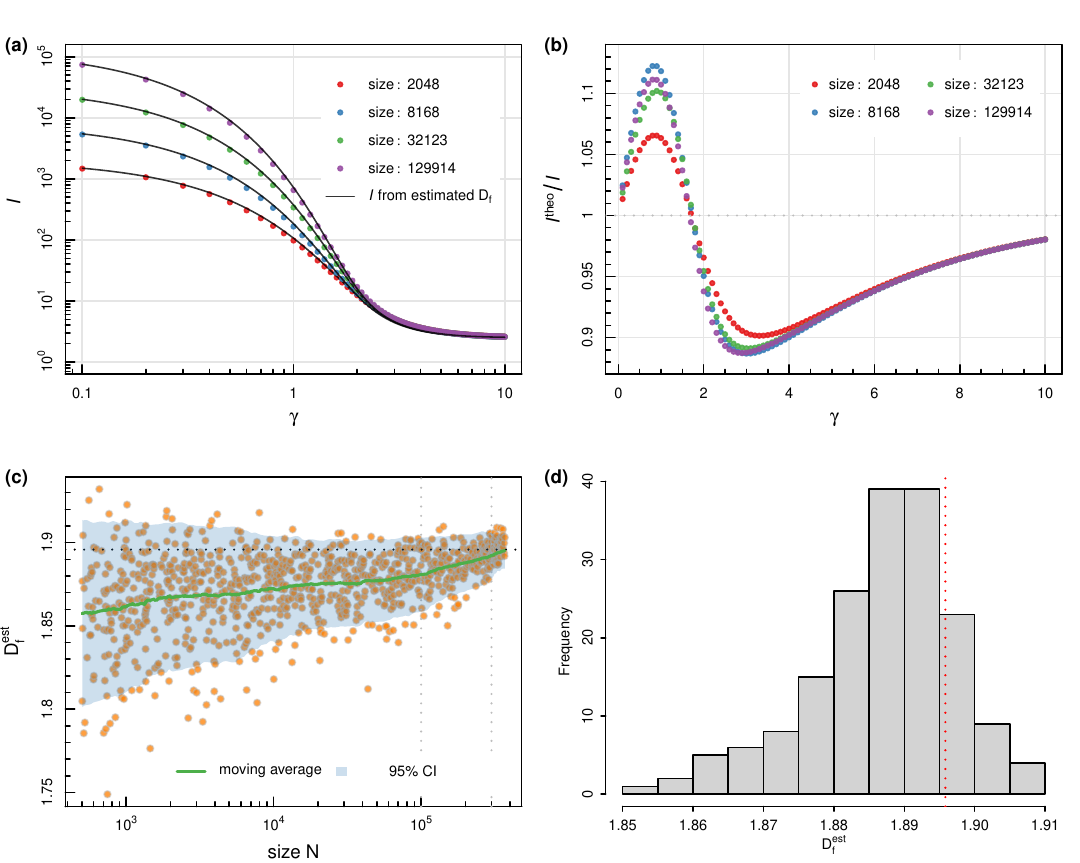}
    \caption{
Estimating the fractal dimension using Eq.~(\ref{eq:final_doublesum}).
(a) $I$ plotted against $\gamma$ for four percolation clusters generated with the Leath algorithm (with different sizes). 
The solid lines denote fitted curves Eq.~(\ref{eq:final_doublesum}) by estimating the fractal dimension $D_f$ as tuned parameter with known $N$, $N_0$, $N_1$, $r_1$, and $r_2$, where $N_1$ is calculated numerically via its definition as the average number of nearest neighbors at the distance $r_1$ that each grid has, and $N_0$ is calculated via Eq.~(\ref{eq:N0N1}). 
(b) The quotient of fitted and numerical values is plotted versus $\gamma$.    
(c) The estimated fractal dimensions for 1,000 percolation clusters generated with the Leath algorithm are plotted against the corresponding structure size.
The solid green line denotes the moving average with a sliding window of 100 samples, and the light blue area indicates the 95\,\% confidence interval.
(d) Histogram of estimated fractal dimensions for percolation clusters with sizes between 100,000 and 300,000, the range is indicated in panel~(c) with two vertical dotted lines.
The horizontal dotted line in panel~(c) and the vertical red dotted line in panel~(d) represent the theoretical fractal dimension of the percolation clusters: $D_f= \frac{91}{48}\approx 1.896$.
    }
    \label{fig:Leat_fitting}
\end{center}
\end{figure*}

Figure~\ref{fig:Leat_fitting}(a)+(b) shows the curves for an example analogous to Fig.~\ref{fig:regular_fractal_fitting}.
The results are equivalent (deviations of approximately 10\,\% in both directions), and from this, we can conclude that our approach is also effective for random fractals.

Repeating the procedure for 1,000 realizations of percolation clusters we obtain a distribution of fractal dimension estimates that we can compare with the theoretical value $\approx$1.9.
In Fig.~\ref{fig:Leat_fitting}(c), the estimates are plotted as a function of the size~$N$.
For small sizes, the average is below the theoretical value (the difference is roughly 0.05 for $N=10^3$), and the spread of estimates is considerable.
The finite clusters are fractal only on scales below their size. 
But the theoretical $D_f$ holds for the ``infinite'' cluster only.

For large sizes, the average approaches the theoretical value and the spread shrinks (95\,\% confidence interval is approximately $\pm0.025$ for $N=10^5$).
In Fig.~\ref{fig:Leat_fitting}(d) the distribution of estimates is shown for a range of sizes.
On the one hand, the estimates are spread around the theoretical value.
On the other hand, the distribution is skewed towards smaller values -- which is due to including small cluster sizes.

Accordingly, Eq.~(\ref{eq:final_doublesum}) can be used as part of a method to estimate the fractal dimension of a spatially distributed structure, but it needs to be sufficiently large to permit a reliable estimate.
A method to estimate $D_f$ could consist of the following steps.
\begin{enumerate}
\item Determine the global information, i.e.\ $N$, $N_0$, $N_1$, $r_1$, and $r_2$;
\item numerically calculate $I$ via Eq.~(\ref{eq:doublesum}) for a range of $\gamma$ values; and
\item fit the numerical $I$ as a function of $\gamma$ using $I^{\rm{theo}}$ from Eq.~(\ref{eq:final_doublesum}) with $D_f$ as a free parameter.
\end{enumerate}
The resulting $D_f$ value represents the estimate of the method.

\section{Summary \& Discussion}
In summary, we treat the problem of estimating the total interactions between any pair of cells of a (fractal) structure.
Calculating the interactions -- defined as the Euclidean distance raised to a power (a kind of gravity model) -- can be computationally expensive, and we derive an analytical expression (involving the fractal dimension).
We provide an analytical derivation and discuss its mathematical properties, including the relation to the Grassberger-Procaccia algorithm.
The idealized solution to this problem, Eqs.~(\ref{eq_result_anali}), works in the long-range regime but fails in the short-range regime.
Discreteness, due to the rasterized nature of the considered structures, inhibits the continuum approximation, which represents an assumption made in the derivation.
Thus, we refine the analytical expression, Eq.~(\ref{eq:final_doublesum}), by explicitly treating the closest neighbors and find considerable improvement when validated numerically.
Last, we explore how this expression can be used to estimate the fractal dimension when unknown.

Our expressions Eqs.~(\ref{eq_result_anali}) and~(\ref{eq:final_doublesum}) represent a simple shortcut so that one does not need to calculate the interactions between any pair of cells numerically, which can be a computationally expensive task.
The interactions between any pair of cells are of interest in many situations, such as social contacts in cities \cite{ribeirocity2017} or urban climate \cite{LiSKR2020}.
We expect new insight into these and many other scientific settings from our theoretical expressions.

To derive our refined expression Eq.~(\ref{eq:final_doublesum}), we have initially separated the interaction of the closest neighbors.
This can certainly be extended to the second and third closest neighbors, and we expect a respective improvement.
However, for the sake of simplicity, we restrict our treatment to Eq.~(\ref{eq:final_doublesum}) as it is already more detailed than Eq.~(\ref{eq_result_anali}).
It is valid to say that further expansion will bring additional parameters and, consequently, increase the difficulty of treating the problem analytically.
The optimal choice of model also depends on the amount of available data for fitting the parameters.

One direct application of the theoretical results presented here is the calculation of the \emph{average distance} $\langle r \rangle$ between cells, which is a specific case of our derivation ($\gamma=-1$, see Sec.~\ref{ssec:exploring}). 
For example, in the context of cities, it has been stated that ``the average distance between any two points inside a circle with area $a$ is given by $128\sqrt{a}/(45\pi)$'' \cite{PrietoCurielPA2023}, which implies $\langle r \rangle \sim a^{1/2}$. 
This consideration was also done in \cite{Xu2020c}.
Our expressions represent a generalization as the circle is just a particular case with $D_f=2$.
For $\gamma=-1$, we obtain $I/N\sim N^{1/D_f}$, which for $D_f=2$ results in $I/N\sim N^{1/2}$, consistent with above statement, given that the area scales with $N$ in our context.
In the case of a city in the form of a line \cite{PrietoCurielK2023} we have $D_f=1$ and the average distance ($\gamma=-1$) scales as $I/N\sim N$.
Accordingly, $(I/N)_{D_f = 2} < (I/N)_{D_f = 1}$ (for $N>1$) - i.e. the average distance in a plain is smaller than the average distance in a line - suggesting that cities expand into the plain if they can.

Models that assume that the interaction decays with the distance, 
as the one treated here, Eq.~(\ref{Eq_power_law_decay}), are also known as \emph{gravity models}.
Initially developed in physics, gravity found numerous applications beyond its traditional domain. 
For instance, for more than 170 years \cite{Philbrick1973}, gravity models have been employed in urban planning to describe and predict the flow of people, goods, and services between different locations within a city or region \cite{Haynes1985,Barthelemy2019,BARTHELEMYbook}.
They provide a versatile framework for analyzing and predicting interactions between entities, enabling researchers and practitioners to make informed decisions and design effective strategies. 
In biology, gravity models have been used to study animal migration patterns 
\cite{LeungBL2006}, the spread of diseases \cite{truscott2012evaluating,barrios2012using},
and to model the competitive and cooperative interactions between individuals \cite{SantosRM2015,Ribeiro2015b,Cabella2012a,ribeiro_tumor2017}.

As it was discussed in previous sections, the decay exponent $\gamma$ controls the range of interactions between the cells, and according to the value of this parameter, one has short- or long-range interactions, leading to different macroscopic behaviors.
This character of the model studied here can be used to describe or even to establish analogies to some phenomena in nature. 
For instance, examples of short-range interactions in physics include the \emph{van der Waals force}
\cite{Israelachvili1974TheNO}, which arise from %
electron distribution around atoms or molecules, and the \emph{strong nuclear force}
\cite{Lacroix2010IntroductionS}, which binds protons and neutrons within atomic nuclei.
Another example of a short-range interaction is the chemical communication in ant communities through pheromones, an excreted chemical substance -- a signal -- that triggers a social response by other ants \cite{JacksonM1993}. 
Pheromones are very volatile molecules which diffuse very fast so the communication between the ants is very restricted to the locality of the signal \cite{RobinsonGJHR2008}.

In contrast, the spread of information, ideas, or behaviour can occur through long-range interactions \cite{ribeirocity2017,LengDMP2023}, affecting communities and societies as a whole. 
Other examples of long-range interaction can be found in social networks, where individuals can be connected to others who are geographically distant \cite{Barthelemy2011,Piva2021}, and in physics, as is the case of self-gravitating systems, wave-particle interacting systems, and non-neutral plasmas \cite{Dauxois2002,Bouchet2010}.
In conclusion, these examples, among many others, can be modeled and comprehended within the theoretical framework presented in this study.

It is important to acknowledge and address potential caveats before concluding the discussion.
The attentive reader will probably observe that while long-range properties work fine, the short-range counterpart is trouble.
Many of the parameters that need to be treated unavoidably are related to the properties at short scales.
Since these short scales are affected by discreteness, it is difficult to treat them accurately.
An example is $N_0$ as introduced in Eq.~(\ref{eq_fractal}).
According to the definition, it is the average number of cells inside a circle of radius $r=1$, as $N(r)=N_0$ for $r=1$.
But what if the smallest distance between any two occupied cells of the considered structure is larger than $1$?
Then $N_0=0$.
Certainly, this situation implies a deviation from Eq.~(\ref{eq_fractal}) and the power-law relation holds true only for asymptotic larger scales.
Similar problems also affect other quantities that our theoretical expressions involve.
On the one hand, this makes it difficult to say what they actually represent.
On the other hand, a lot of the content of our paper is dedicated to working around these problems.
In some cases, these quantities are simply used as fitting parameters.

We also propose to use our theoretical expression(s) -- in combination with numerically calculating the interactions between all pairs and for various exponents -- as a method to estimate the fractal dimension.
At this point, the idea has only been shown exemplarily (\emph{proof of concept}).
Further research is necessary to assess the potential of such a method.
First, we only investigated one type of fractal, and observed deviations could be due to its generation, i.e.\ the theoretical fractal dimension could only be achieved for asymptotically large percolation clusters. 
Second, it is likely that more elaborated fitting could provide better estimates (e.g.\ maximum likelihood).
Third, it will be interesting to compare the performance of the proposed method with established ones (e.g.\ box-counting/covering or sandbox methods \cite{fractal-disor-book,book-fractals1994} involve a set of details that affect the estimate).

Another way to extend our work could be to analyze structures where the cells are more than binary.
The work in hand is restricted to grids with cells that are either empty or occupied.
In many real-world situations, the cells carry some sort of weight, and it could be relevant to derive analogous expressions for such more complex systems.

\begin{acknowledgments}
We thank J.W.\ Kantelhardt for useful comments
as well as A.\ Martinez and G.\ Nakamura for insights into the generalized functions and collapsed curve.
Y.\ Li and D. Rybski would like to thank German Research Foundation (DFG) for funding this research within the \emph{Urban Percolations} project (451083179).
D.\ Rybski is grateful to the Leibniz Association (project CriticaL) for financially supporting our research.
D.\ Rybski thanks the Alexander von Humboldt Foundation for financial support under the Feodor Lynen Fellowship.
F.\ L.\ Ribeiro thanks CNPq (grant numbers 403139/2021-0 and 424686/2021-0) and Fapemig (grant number APQ-00829-21) for financial support.

\end{acknowledgments}



\appendix

\section{Exploring the analytical result (continuation)}
\label{app:exploring}

In this Appendix section, we continue exploring the analytical result Eq.(\ref{eq_result_anali}) to predict the interaction field for specific $\gamma$ values.
This exploration complements the findings discussed in Sec.~\ref{ssec:exploring}.

\begin{itemize}
\item $\gamma = 0$ (physical distance does not matter) yields 
\begin{equation}
    I^{\rm{theo}} = N - r_1^{D_f} \, .
\end{equation}
Moreover, $r_1 =1$ implies $I^{\rm{theo}} = N - 1$, which is expected when each cell interacts with every other cell except with itself.

\item $\gamma = D-1$, where $D$ is the \emph{Euclidean dimension} of the medium in which the fractal structure is embedded. 
It corresponds to the general form of Newton's law of gravitation
in Eq.~(\ref{Eq_power_law_decay}).
The analytical solution Eq.~(\ref{eq_result_anali})  for this context yields
\begin{equation}
I^{\rm{theo}} = \frac{N_0 D_f}{(D_f - D +1)} \left[ \left( \frac{N}{N_0} \right)^{1- \frac{(D-1)}{D_f}} -  r_1^{D_f - D +1} \right]
\, .
\end{equation}
For our specific study, in which the structures are embedded  in a two-dimensional medium, one has $D=2$, which yields
\begin{equation}
I^{\rm{theo}} = \frac{N_0}{(1 - \frac{1}{D_f})} \left[ \left( \frac{N}{N_0} \right)^{1 - \frac{1}{D_f}} -  r_1^{D_f -1} \right]
\, .
\end{equation}

\item $\gamma = - k $, where
$k$ is a natural number ($k=1,2,3,\dots$). 
In this case, the analytical solution Eq.~(\ref{eq_result_anali}) yields the moments of the distance distribution of the cells.
Therefore, we use the integral in Eq.~(\ref{eq_aproxx}) and $p(r)dr = dN(r)/N$, where $p(r)dr$ is the probability of finding a cell between $r$ and $r+dr$.
In this way, for $\gamma =-k$, Eq.~(\ref{eq_result_anali}) becomes
\begin{equation}
I^{\rm{theo}} = N \int r^{k} p(r)dr = N \langle r^k \rangle    \, ,
\end{equation}
where we identify the relation between the mean interaction field and the  $k$-th moment of the distance distribution of cells:  %
$\langle r^k \rangle~=~\int r^{k} p(r)dr$.
In this way, the result Eq.~(\ref{eq_result_anali}) allows us to get directly the $k$-th moment of the distance distribution of a fractal spatial arranged population. 
More specifically, using the analytical solution Eq.~(\ref{eq_result_anali})  one has
\begin{equation}
 \langle r^k \rangle  = \frac{N_0}{N (1 + \frac{k}{D_f})} \left[ \left( \frac{N}{N_0} \right)^{1+ \frac{k}{D_f}} -  r_1^{D_f +k} \right]
\, .
\end{equation}

\item  $\gamma = -1$ represents a particular situation of the previously discussed case.
The interaction field is directly related to the average distance between cells by the form $I^{\rm{theo}} = N \langle r \rangle$.
Consequently, from the result Eq.~(\ref{eq_result_anali}) it is possible to directly obtain the average distance between cells of fractal structure, given by
\begin{equation}
\langle r \rangle  = 
\frac{N_0}{N(1 + \frac{1}{D_f})} \left[ \left( \frac{N}{N_0} \right)^{1 + \frac{1}{D_f}} -  r_1^{D_f +1} \right]
\, .
\end{equation}
It is interesting to note that for $ N \gg N_0 $ this leads to  
\begin{equation}
  \langle r \rangle \sim N^{\frac{1}{D_f}} \, .
\end{equation}

\end{itemize}

\begin{figure*}
	\begin{center}
	\includegraphics[width=0.75\textwidth]{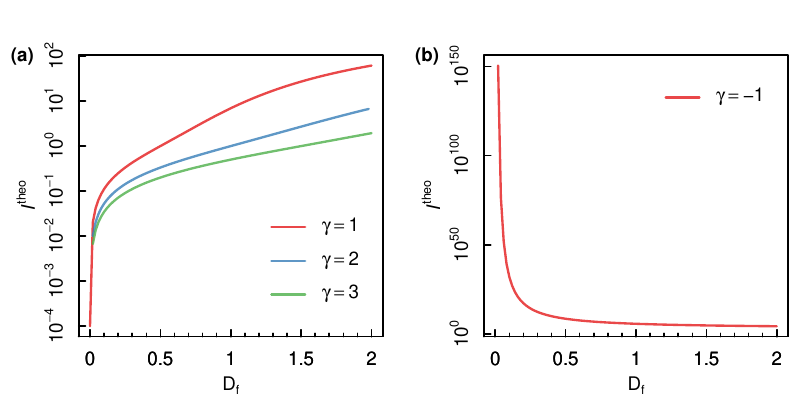}
    \caption{
\label{fig:I_Df_plot}
The $I$ against fractal dimension $D_f$ for Eq.~\ref{eq_result_anali} with $N=1000$, $N_0 = 1$ and $r_1 = 1$. Coloured lines are from different $\gamma$ values.
}
\end{center}
\end{figure*}

Last, in Fig.~\ref{fig:I_Df_plot} we illustrate how $I^{\rm{theo}}$ depends on the fractal dimension.
Figure~\ref{fig:I_Df_plot}(a) depicts curves for $\gamma>0$ (interactions decrease with the distance), one can see that in this case $I^{\rm{theo}}$ increases with $D_f$, i.e.\ more compact structures lead to more interactions.
Figure~\ref{fig:I_Df_plot}(b) depicts a curve for $\gamma=-1$, the total distances between any pair of cells.
In this case, $I^{\rm{theo}}$ decreases with $D_f$, i.e.\ more compact structures lead to smaller total distances -- which is intuitively expected.

\section{Writing Eq.~(\ref{eq_result_anali}) in terms of generalized logarithm}
\label{sec:genlog}

The mathematical form of the analytical solution Eq.~(\ref{eq_result_anali}) suggests similarities with the definition of the generalized logarithm
\cite{Ribeiro2015b,Cabella2012a}. 

The natural logarithm $\ln(x)$ can be defined as the area below the hyperbole $f(t) = 1/t$, from $t=1$ to $t=x$, that is  
\begin{equation}
 \ln(x) = \int_1^x \frac{1}{t} dt
 \, .
\end{equation}
Similarly, we can define the \emph{generalized logarithm} as the area below the \emph{generalized hyperbole}, defined as $1/t^{1-q}$, where $q$ is the \emph{generalization parameter}, and $q=0$ recovers the hyperbole. 
If we call $\ln_q(x)$ the generalized logarithm of $x$, defined in the interval $t \in [1,x]$, we can formally write
\begin{equation}\label{Eq_gene_log}
\ln_{q}(x)=  \int_1^x \frac{dt}{t^{1-q}} =
\left\{ \begin{array}{ll}
\frac{x^q -1}{q} & \textrm{for } q \ne 0 \\
\ln(x)  & \textrm{for } q \to 0
\end{array} \right. \, .
\end{equation}
Note that, in such a notation, $q$ should not be misunderstood as the base of the logarithm.

Using the generalized logarithm definition Eq.~(\ref{Eq_gene_log}), and assuming $r_1=1$, the solution Eq.~(\ref{eq_result_anali}) can be simplified to
\begin{equation}\label{eq_result_anali_generalized}
I^{\rm{theo}} = N_0 \ln_q \Big(\frac{N}{N_0}\Big) \, ,
\end{equation}
where 
\begin{equation}
    q = 1 - \frac{\gamma}{D_f} \, .
\end{equation}
Now it is easy to verify that in the case of $\gamma = D_f$ (and consequently $q = 0$) the interaction field $I^{\rm{theo}}$ growths logarithmically with $N$, that is $I^{\rm{theo}} = N_0 \ln \Big(\frac{N}{N_0}\Big)$.

\section{Details on relation to Grassberger-Procaccia algorithm}
\label{sec:detailsGP}

In this present work, the interaction strength between two cells, namely $I_{ij}$, is studied using a specific power-law decay with the distance, defined in Eq.~(\ref{Eq_power_law_decay}). 
But of course, we could proceed with our studies using other functions to model this pair interaction, for instance using 
\begin{equation}\label{eq_theta}
 I_{ij} \equiv \Theta(r_{ij} - a) \, ,  
\end{equation}
where $\Theta(r_{ij} - a)$ is the Heaviside step function. 
It represents the case in which there is an interaction between $i$ and $j$ only if they are separated by a distance smaller than $a$.
This scenario is particularly interesting because it corresponds to a mean interaction field that has the same mathematical structure as the \textit{Grassberger-Procaccia} (GP) \textit{algorithm}. 
As part of the algorithm, given $N$ spatially distributed cells, the quantity 
\begin{equation}
\label{eq:GPdef2}
C(a) = \frac{2}{N(N-1)} \sum_i^N \sum_{j\ne i} \Theta{(a-r_{i_j})}
\end{equation}
is defined.
As it was shown in the original paper \cite{GrassbergerP1983prl}, this quantity scales as 
\begin{equation}
C(a) \sim a^{D_c}
\end{equation}
where $D_c$ is the so-called \emph{correlation fractal dimension}.

As an exercise, we can solve Eq.~(\ref{eq:GPdef2}) using the approach presented in the main paper. 
To do this, we can employ the continuum approximation 
\begin{equation}\label{eq_aproxx2}
I_i = \sum_{j\ne i}^N \theta{(a-r_{i_j})}
 \quad \to  \quad  I_i^{\rm{theo}} \equiv \int \theta{(a-r)} dN(r)
\end{equation}
and suppose a fractal structure with dimension $D_f$, i.e.\ $dN(r)$ obeying Eq.~(\ref{eq:dNDf}) as before. Consequently 
\begin{equation}
\begin{aligned}
I_i^{\rm{theo}}  & =  \int_{r_1}^{R_{max}} \theta{(a-r)} N_0 D_f r^{D_f -1} dr \\
    & = N_0 D_f\int_{r_1}^{a} \ r^{D_f -1} dr \, ,
\end{aligned}
\end{equation}
which results in 
\begin{equation}\label{eq_aproxx4}
I_i^{\rm{theo}} = N_0 a^{D_f}
\, ,
\end{equation}
for $a>r_1$.
This makes sense because it  corresponds exactly to the number of cells inside a circle with radius $a$ centred in $i$ (given Eq.~(\ref{eq_fractal}) holds).

Combining Eqs.~(\ref{eq_aproxx4},\ref{eq_aproxx2},\ref{eq_fractal}) one gets
\begin{equation}
    C(a) = \frac{2}{N-1} N_0 a^{D_f} 
\end{equation}
and for fixed $N$ it is $C(a) \sim  a^{D_f}$.
This means that both approaches, by Grassberger \& Procaccia and following our derivation, lead to analogous power-laws, restricted to monofractal structures, for which correlation and fractal dimensions are the same, $D_c = D_f$.
It shows that our approach agrees with the results of the GP algorithm.

Continuing such analyses, we could try to identify under which circumstance our approach, considering $I_{ij} = 1/r_{ij}^{\gamma}$ depending on the parameter $\gamma$, is similar to the model 
leading to Eq.~(\ref{eq_theta}), which depends on the parameter $a$.
In fact, they are similar when both present the same value of the interaction field $I_i = \sum_{j \ne i} I_{ij}$. 
Specifically, restricting our approach to the long-range interaction regime, one gets $I_i \sim N^{1-\gamma/D_f}$, conform with Eq.~(\ref{Eq_long-range}), and for the model Eq.~(\ref{eq_theta}) one gets  $I_i \sim a^{D_f}$.
Equaling these two results yields 
\begin{equation}
    \gamma =   D_f - D_f^2\frac{\ln a}{\ln N} + {\rm cte} \cdot \frac{D_f}{\ln N} \, ,
\end{equation}
which represents the situation in which our approach recovers the GP algorithm.

\section{Overview of patterns}
\begin{figure*}
	\begin{center}
	\includegraphics[width=0.9\textwidth]{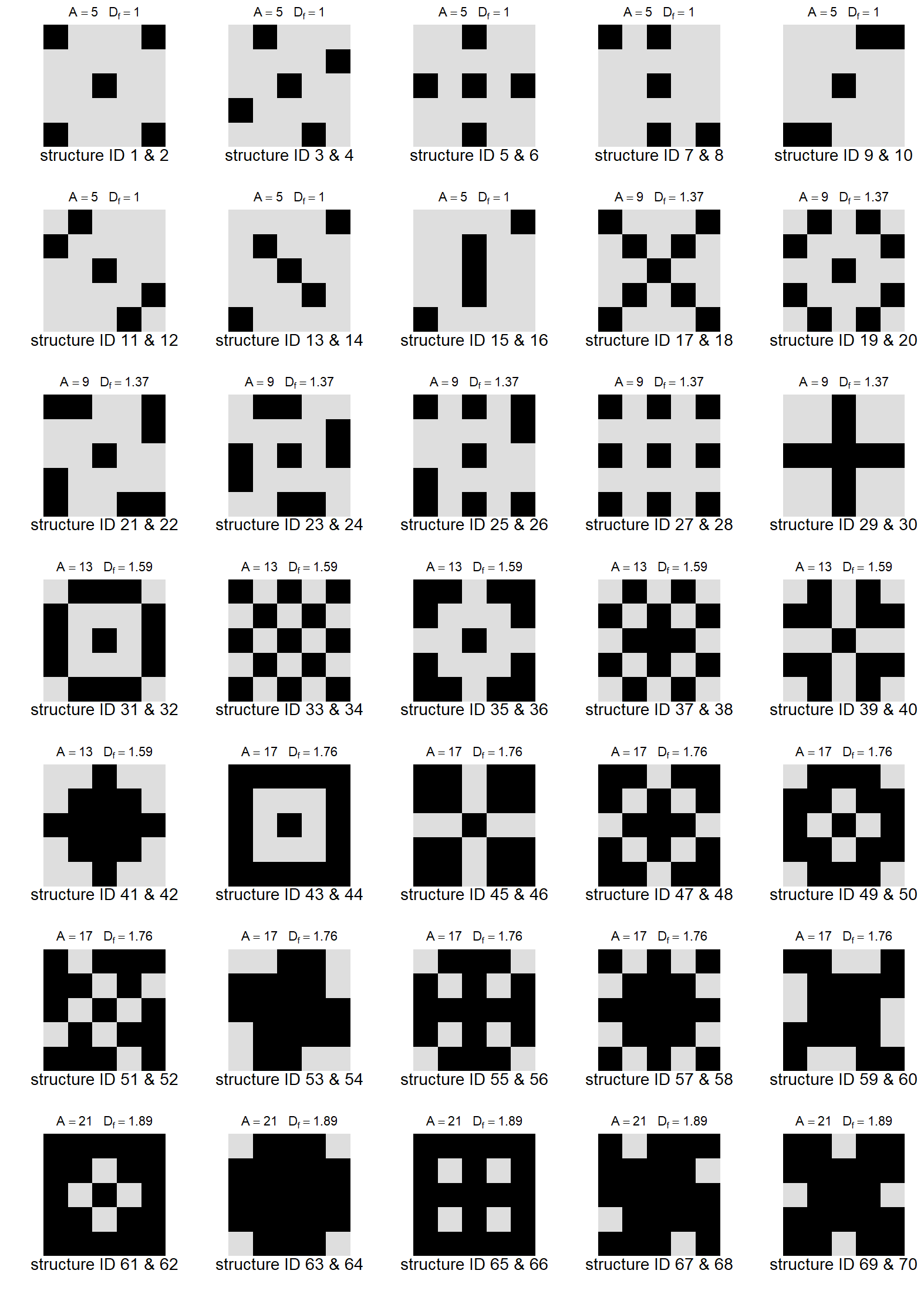}
    \caption{
    Template patterns for creating regular fractal structures. We create 2 structures from each base pattern, at third iteration and fourth iteration, respectively. Creating structures from growing the base pattern is illustrated in Fig.~\ref{fig:regular_fractal_growth}}.
    \label{fig:base_structure}
\end{center}
\end{figure*}

\begin{figure*}
	\begin{center}	\includegraphics[width=0.9\textwidth]{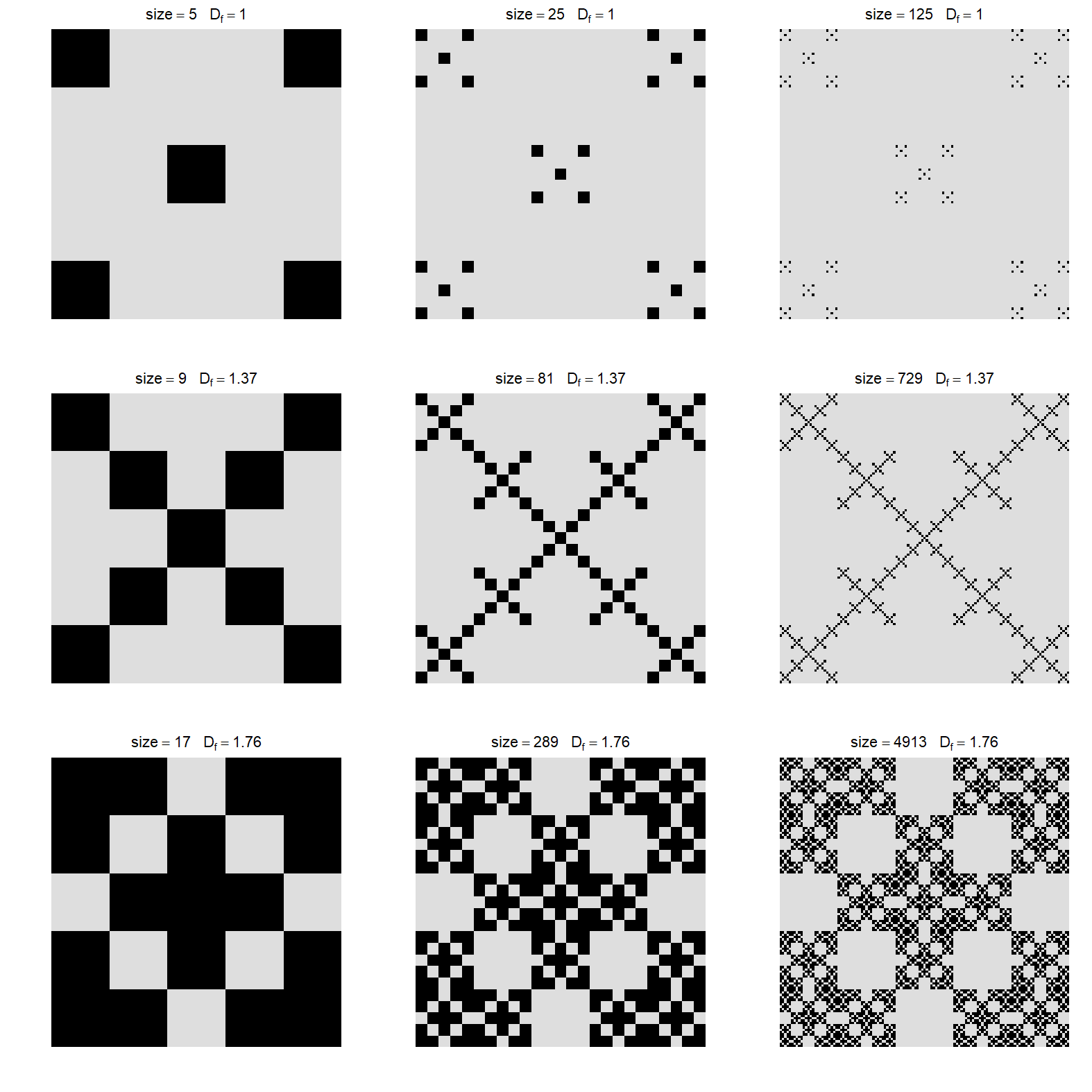}
    \caption{Illustration of the growth of the structure from base structure to the first interaction, and the second iteration. Note the patterns actually used in the analysis are from third and fourth iterations and they are now shown in this figure due to their too large sizes.}
    \label{fig:regular_fractal_growth}
\end{center}
\end{figure*}

\begin{figure*}
	\begin{center}	
    \includegraphics[width=0.9\textwidth]{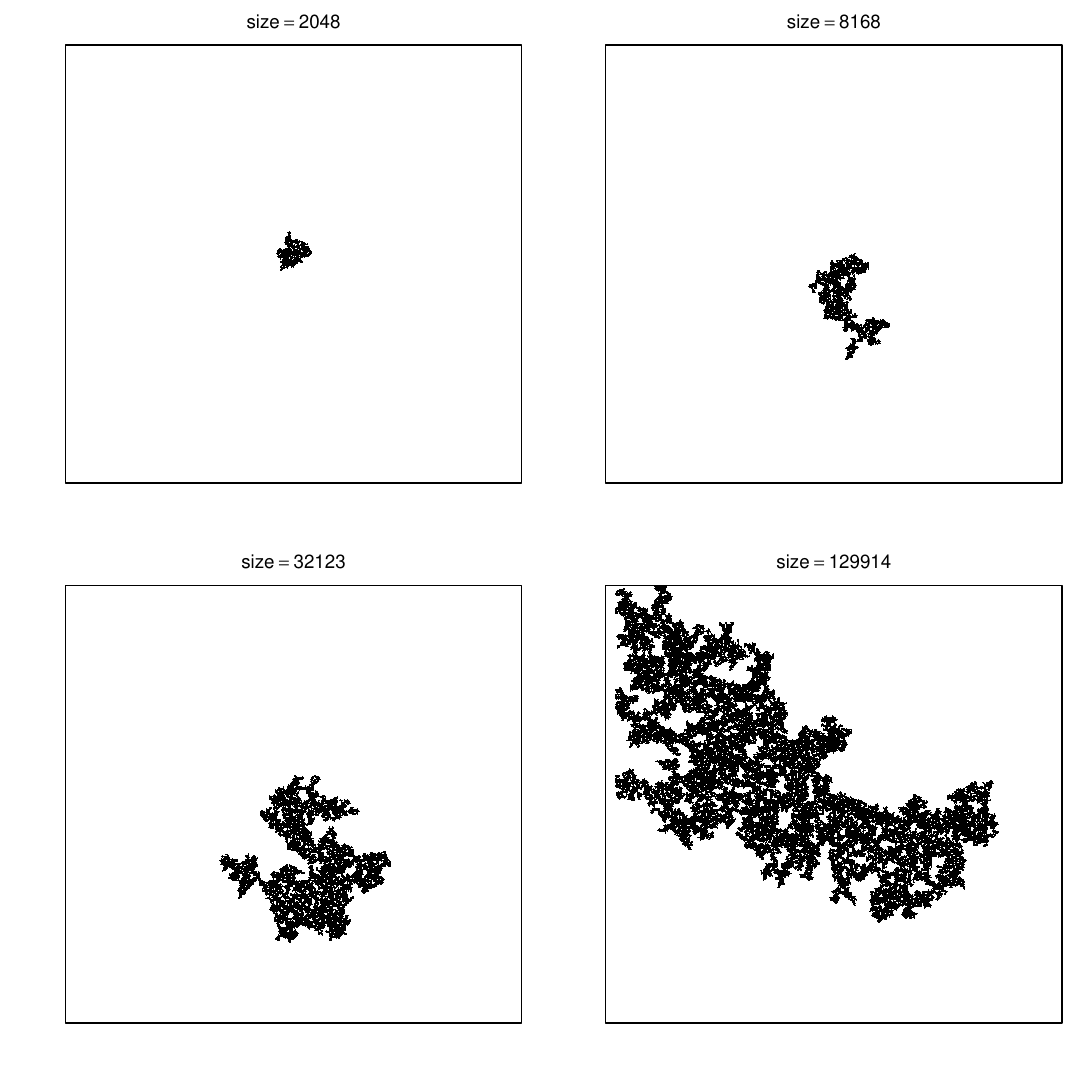}
    \caption{Visualization of 4 percolation clusters generated with the Leath algorithm, used in Fig.~\ref{fig:Leat_fitting}a.}
    \label{fig:Leathviz}
\end{center}
\end{figure*}

See Figs.~\ref{fig:base_structure}, \ref{fig:regular_fractal_growth}, and \ref{fig:Leathviz}.

\end{document}